\documentclass[10pt,conference]{IEEEtran}
\IEEEoverridecommandlockouts
\usepackage{cite}
\usepackage{amsmath,amssymb,amsfonts}
\usepackage{algorithmic}
\usepackage{graphicx}
\usepackage{textcomp}
\usepackage{xcolor}
\usepackage{booktabs}
\usepackage{colortbl}
\usepackage{array}
\usepackage{tikz}
\usepackage{xspace}
\usepackage{fontawesome5}  %
\usepackage{pifont}        %
\usepackage{url}           %
\usetikzlibrary{arrows.meta,positioning,calc}
\def\BibTeX{{\rm B\kern-.05em{\sc i\kern-.025em b}\kern-.08em
T\kern-.1667em\lower.7ex\hbox{E}\kern-.125emX}}

\newcommand{\ourSystemName}{CodeMeepo}
\newcommand{\ourSystem}{\textsc{\ourSystemName}\xspace}

\newcommand{\cat}[1]{\smallskip\par\noindent\textbf{#1}\hspace{0.5em}\ignorespaces}

\newcommand{\stepn}[1]{\ding{\numexpr201+#1\relax}}
\IEEEoverridecommandlockouts
\begin{document}
\title{\ourSystemName: A Multi-Agent Workflow for\\Repository-Level Code Migration and Optimization}

\author{
\IEEEauthorblockN{Farzad Habibi\IEEEauthorrefmark{1}\IEEEauthorrefmark{4}, Changrong Wu\IEEEauthorrefmark{2}, Jinzhou Jiang\IEEEauthorrefmark{2}, and Guohan Lu\IEEEauthorrefmark{3}\IEEEauthorrefmark{4}}
\IEEEauthorblockA{\IEEEauthorrefmark{1}\textit{University of California, Irvine}, \IEEEauthorrefmark{2}\textit{Microsoft}, \IEEEauthorrefmark{3}\textit{OpenAI}}
\IEEEauthorblockA{\IEEEauthorrefmark{1}habibif@uci.edu, \IEEEauthorrefmark{2}\{changrongwu, riffjiang\}@microsoft.com, \IEEEauthorrefmark{3}lgh@openai.com}
\IEEEauthorblockA{\IEEEauthorrefmark{4}\textit{All contributions were made while at Microsoft.}}
}
\maketitle

\begin{abstract}
Repository-level code migration rewrites an entire project in a new programming language while preserving its observable behavior. This task is challenging because behavior spans interdependent modules, dependencies and build infrastructure must be adapted to the target ecosystem, and the source code often serves as the only specification~\cite{googlemigration,locksmith,cruststudy}.

Large language model agents can now attempt these migrations end to end~\cite{alphatrans,recodeagent}, but three challenges remain. First, an agent that writes both the translation and its acceptance tests may overlook the same requirements in both, leaving missing or incorrect behavior undetected. Second, large repositories can exceed a model's context window, and dividing the work into smaller tasks does not ensure that all required behavior is migrated. Third, even behaviorally correct translations may be slower than the original implementation. 

We present \ourSystem, a multi-agent pipeline designed for complete and efficient migration of large repositories. Each milestone is evaluated against a fixed external oracle that the agents neither author nor modify. A separate parity assessment compares source and target components within the migration scope and adds identified gaps back into the plan as required work. Ordered milestones limit the scope of each task, while cumulative regression checks guard against breaking previously migrated behavior. A benchmark-guided optimization phase addresses performance, followed by full-suite conformance testing and repair.
Eight specialized agent roles communicate through file artifacts, while a deterministic state machine governs repair, deferral, scope changes, and termination.

On nine subjects in three language pairs, \ourSystem achieves the best pass rate on every subject and passes 18\% more tests than a state-of-the-art multi-agent baseline using the same model. On a production daemon, its optimization phase then improves half of the benchmarks, by 25\% on average.
\end{abstract}


\section{Introduction}
\label{sec:introduction}

Programming-language migration allows organizations to maintain software when its original implementation no longer meets their security, interoperability, or maintainability requirements~\cite{googlemigration}. Organizations move business applications from COBOL to Java~\cite{bhardwaj2026cobol,chakaravarthy2026long}, port components between general-purpose languages, and replace C with memory-safe Rust~\cite{cruststudy}. Such changes are not simply syntactic translations. They must preserve established behavior while replacing dependencies, interfaces, build procedures, and deployment assumptions. Across language pairs, practitioners report that the dominant cost is not writing the target code but establishing that it behaves like the original~\cite{locksmith}.

Large language models (LLMs) have expanded the range of migration tasks that can be automated. Research has progressed from translating isolated functions~\cite{transcoder,transcoderst} to planning repository-wide edits~\cite{codeplan} and translating interdependent project components~\cite{alphatrans,recodeagent}. Multi-agent approaches distribute the work across specialized roles, while retrieval can supply migration-specific knowledge~\cite{comra2026}. These developments make repository-level migration a promising application of agentic software engineering, and raise the question this paper addresses: \emph{how can agents produce a complete and efficient port of a large repository without human intervention?}

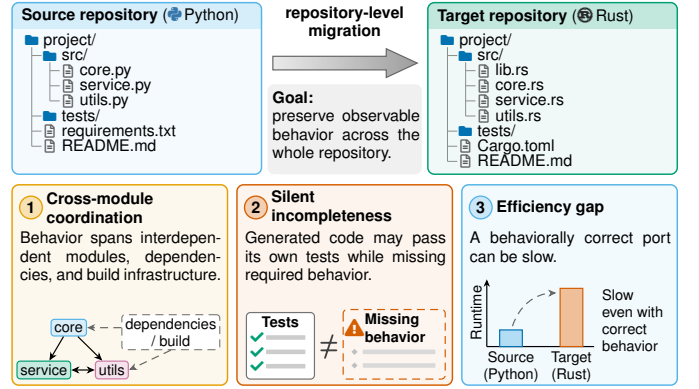
\begin{figure}[t]
\centering
\definecolor{oiBlue}{HTML}{0072B2}%
\definecolor{oiSky}{HTML}{56B4E9}%
\definecolor{oiOrange}{HTML}{E69F00}%
\definecolor{oiVerm}{HTML}{D55E00}%
\definecolor{oiGreen}{HTML}{009E73}%
\definecolor{oiPurple}{HTML}{CC79A7}%
\definecolor{pyBlue}{HTML}{3776AB}%
\begin{tikzpicture}[
  x=1cm, y=1cm,
  font=\sffamily\fontsize{6}{7}\selectfont,
  small/.style={font=\sffamily\fontsize{5.5}{6.5}\selectfont},
  smallbf/.style={font=\sffamily\fontsize{5.5}{6.5}\selectfont\bfseries},
  guide/.style={draw=black!45, line width=0.35pt},
  mod/.style={small, draw=#1, fill=#1!22, line width=0.45pt, rounded corners=1.2pt,
    inner sep=1.3pt, minimum height=0.22cm},
  dep/.style={-{Stealth[length=1.1mm,width=0.95mm]}, line width=0.45pt},
  both/.style={{Stealth[length=1.1mm,width=0.95mm]}-{Stealth[length=1.1mm,width=0.95mm]},
    line width=0.45pt},
  ext/.style={dep, densely dashed, black!60},
  badge/.style={circle, draw=#1, fill=#1!30, line width=0.5pt, inner sep=0pt,
    minimum size=0.3cm, font=\sffamily\fontsize{6}{7}\selectfont\bfseries},
  card/.style={draw=#1, fill=#1!7, line width=0.55pt, rounded corners=2.5pt},
  ctitle/.style={anchor=west, align=left, inner sep=0pt,
    font=\sffamily\fontsize{6}{7}\selectfont\bfseries},
  just/.style={align=justify,
    execute at begin node={\tolerance=3000\emergencystretch=0.6em\relax}},
  cbody/.style={anchor=north west, text width=2.66cm, just, inner sep=0pt},
]
\def\icfolder{{\color{oiBlue}\fontsize{5}{6}\selectfont\faFolder}}
\def\icfile{{\color{black!65}\fontsize{5}{6}\selectfont\faFile*[regular]}}
\def\titem#1#2#3#4{%
  \node[inner sep=0pt] at (#1,#2) {#3};
  \node[anchor=base west, inner sep=0pt] at ($(#1,#2)+(0.15,-0.06)$) {#4};}
\def\repo#1#2#3#4{%
  \fill[#2!6, rounded corners=2.5pt] (#1,0) rectangle ++(3.3,-2.27);
  \fill[#2!24, rounded corners=2.5pt] (#1,0) rectangle ++(3.3,-0.34);
  \fill[#2!24] (#1,-0.2) rectangle ++(3.3,-0.14);
  \draw[#2, line width=0.55pt, rounded corners=2.5pt] (#1,0) rectangle ++(3.3,-2.27);
  \node[anchor=west, inner sep=0pt] at ($(#1,-0.17)+(0.12,0)$) {\textbf{#3} #4};}

\repo{0}{oiSky}{Source repository}{({\color{pyBlue}\faPython}\,Python)}
\repo{5.5}{oiGreen}{Target repository}{({\color{black!85}\faRust}\,Rust)}

\begin{scope}[shift={(0,0)}]
  \draw[guide] (0.26,-0.59) -- (0.26,-1.9);
  \foreach \y in {-0.7,-1.5,-1.7,-1.9} {\draw[guide] (0.26,\y) -- (0.39,\y);}
  \draw[guide] (0.5,-0.79) -- (0.5,-1.3);
  \foreach \y in {-0.9,-1.1,-1.3} {\draw[guide] (0.5,\y) -- (0.63,\y);}
  \titem{0.26}{-0.5}{\icfolder}{project/}
  \titem{0.50}{-0.7}{\icfolder}{src/}
  \titem{0.74}{-0.9}{\icfile}{core.py}
  \titem{0.74}{-1.1}{\icfile}{service.py}
  \titem{0.74}{-1.3}{\icfile}{utils.py}
  \titem{0.50}{-1.5}{\icfolder}{tests/}
  \titem{0.50}{-1.7}{\icfile}{requirements.txt}
  \titem{0.50}{-1.9}{\icfile}{README.md}
\end{scope}
\begin{scope}[shift={(5.5,0)}]
  \draw[guide] (0.26,-0.59) -- (0.26,-2.1);
  \foreach \y in {-0.7,-1.7,-1.9,-2.1} {\draw[guide] (0.26,\y) -- (0.39,\y);}
  \draw[guide] (0.5,-0.79) -- (0.5,-1.5);
  \foreach \y in {-0.9,-1.1,-1.3,-1.5} {\draw[guide] (0.5,\y) -- (0.63,\y);}
  \titem{0.26}{-0.5}{\icfolder}{project/}
  \titem{0.50}{-0.7}{\icfolder}{src/}
  \titem{0.74}{-0.9}{\icfile}{lib.rs}
  \titem{0.74}{-1.1}{\icfile}{core.rs}
  \titem{0.74}{-1.3}{\icfile}{service.rs}
  \titem{0.74}{-1.5}{\icfile}{utils.rs}
  \titem{0.50}{-1.7}{\icfolder}{tests/}
  \titem{0.50}{-1.9}{\icfile}{Cargo.toml}
  \titem{0.50}{-2.1}{\icfile}{README.md}
\end{scope}

\node[align=center, font=\sffamily\fontsize{6}{7}\selectfont\bfseries] at (4.4,-0.27)
  {repository-level\\migration};
\shade[left color=black!8, right color=black!38, draw=black!60, line width=0.45pt]
  (3.45,-0.71) -- (4.95,-0.71) -- (4.95,-0.61) -- (5.36,-0.8)
  -- (4.95,-0.99) -- (4.95,-0.89) -- (3.45,-0.89) -- cycle;
\fill[black!6, rounded corners=2.5pt] (3.38,-1.08) rectangle (5.42,-2.27);
\node[anchor=north west, text width=1.95cm, just, inner sep=0pt]
  at (3.44,-1.16) {\textbf{Goal:}\\ preserve observable behavior across the
  whole repository.};

\draw[card=oiOrange] (0,-2.41) rectangle ++(2.85,-2.65);
\draw[card=oiVerm]   (2.975,-2.41) rectangle ++(2.85,-2.65);
\draw[card=oiSky]    (5.95,-2.41) rectangle ++(2.85,-2.65);

\node[badge=oiOrange] at (0.25,-2.71) {1};
\node[ctitle] at (0.45,-2.71) {Cross-module\\coordination};
\node[cbody] at (0.1,-3.02) {Behavior spans interdependent modules,
  dependencies, and build infrastructure.};

\node[badge=oiVerm] at (3.225,-2.71) {2};
\node[ctitle] at (3.425,-2.71) {Silent\\incompleteness};
\node[cbody] at (3.075,-3.02) {Generated code may pass its own tests while
  missing required behavior.};

\node[badge=oiSky] at (6.2,-2.71) {3};
\node[ctitle] at (6.4,-2.71) {Efficiency gap};
\node[cbody] at (6.05,-3.02) {A behaviorally correct port can be
  slow.};

\node[mod=oiSky]    (core) at (0.75,-4.3)  {core};
\node[mod=oiGreen]  (svc)  at (0.42,-4.86) {service};
\node[mod=oiPurple] (utl)  at (1.32,-4.86) {utils};
\node[small, draw=black!55, densely dashed, fill=white, rounded corners=1.2pt,
  align=center, inner sep=1.3pt] (ext) at (2.1,-4.36) {dependencies\\/ build};
\draw[dep] (core) -- (svc);
\draw[dep] (core) -- (utl);
\draw[both] (svc) -- (utl);
\draw[ext] (ext.west |- core) -- (core.east);
\draw[ext] (ext.south) -- (utl.east);

\draw[draw=black!55, fill=white, rounded corners=1.5pt, line width=0.45pt]
  (3.1,-4.1) rectangle (4.0,-4.98);
\node[smallbf] at (3.55,-4.22) {Tests};
\foreach \y in {-4.43,-4.62,-4.81} {
  \node[inner sep=0pt, oiGreen] at (3.24,\y) {\fontsize{5}{6}\selectfont\faCheck};
  \fill[black!20, rounded corners=0.3pt] (3.36,\y-0.03) rectangle (3.88,\y+0.03);}
\draw[black!75, line width=0.7pt, line cap=round]
  (4.09,-4.5) -- (4.29,-4.5) (4.09,-4.6) -- (4.29,-4.6) (4.24,-4.4) -- (4.14,-4.7);
\draw[draw=oiVerm, densely dashed, fill=oiVerm!6, rounded corners=1.5pt, line width=0.45pt]
  (4.38,-4.06) rectangle (5.73,-4.98);
\node[inner sep=0pt, oiVerm] at (4.53,-4.3) {\fontsize{6.5}{7}\selectfont\faExclamationTriangle};
\node[smallbf, anchor=west, align=left, inner sep=0pt] at (4.66,-4.3) {Missing\\behavior};
\foreach \y in {-4.62,-4.8} {
  \fill[black!25] (4.53,\y) +(0,0.04) -- +(0.04,0) -- +(0,-0.04) -- +(-0.04,0) -- cycle;
  \fill[black!15, rounded corners=0.3pt] (4.64,\y-0.03) rectangle (5.6,\y+0.03);}

\draw[-{Stealth[length=1mm,width=0.85mm]}, line width=0.45pt] (6.28,-4.55) -- (6.28,-3.66);
\draw[line width=0.45pt] (6.28,-4.55) -- (7.7,-4.55);
\node[small, rotate=90] at (6.14,-4.1) {Runtime};
\draw[oiBlue, fill=oiSky!45, line width=0.45pt] (6.45,-4.55) rectangle (6.75,-4.33);
\draw[oiVerm, fill=oiVerm!40, line width=0.45pt] (7.25,-4.55) rectangle (7.55,-3.78);
\node[small, anchor=north, align=center, inner sep=1pt] at (6.6,-4.57) {Source\\(Python)};
\node[small, anchor=north, align=center, inner sep=1pt] at (7.4,-4.57) {Target\\(Rust)};
\draw[ext] (6.62,-4.28) to[bend left=35] (7.2,-3.84);
\node[small, anchor=north west, align=left, inner sep=0pt] at (7.8,-3.78)
  {Slow\\even with\\correct\\behavior};
\end{tikzpicture}
\caption{Repository-level code migration: task and challenges.}
\label{fig:challenges}
\end{figure}

\cat{Three challenges.}
Figure~\ref{fig:challenges} illustrates the three challenges that such a port faces. The first challenge is \emph{scale}. Repository-level code migration requires extensive analysis and planning to coordinate changes across components and dependencies. For large repositories, this work cannot fit within a single model context, and model accuracy degrades as the relevant material is buried in a longer one~\cite{lostinmiddle}. The work must therefore be divided into smaller tasks, each with its own plan, whether by call-graph order~\cite{alphatrans}, dependency-aware planning~\cite{codeplan}, or bounded code chunks~\cite{chakaravarthy2026long}. Decomposition makes tasks more manageable, but introduces a coordination challenge: later changes must not break previously migrated behavior, and the system must track and revisit tasks that remain unresolved.

The second challenge is \emph{self-validation}. When the same agentic process writes both the target implementation and its acceptance tests, both may reflect the same misunderstanding of the source. Generated oracles tend to record the behavior a program exhibits rather than the behavior it should exhibit, so they can encode a defect as though it were the specification~\cite{testoracles}. Passing these tests can therefore create false confidence in an incorrect or incomplete translation, and this is compounded by the tendency of language models to produce code that compiles but is semantically wrong~\cite{lostintranslation}. For example, a port may compile and pass its generated tests while omitting a required interface or an error-handling path. We call a translation that appears complete but lacks required behavior a \emph{silently unfinished port}. The problem is not simply limited test coverage; for generated suites, coverage and mutation scores are themselves unreliable indicators of fault-detection ability~\cite{coveragecorrelate}. The deeper issue is that the implementation and the tests used to accept it may share the same blind spots. Independent verification tests, source-derived checks, and explicit coverage accounting address different parts of this concern.

The third challenge is \emph{efficiency}. Behavioral compatibility does not establish that a translated application has acceptable runtime or resource use, and generated code is frequently far less efficient than competent human implementations~\cite{effibench,coffe}. A maintainable replacement can still be unsuitable if an inefficient data representation or synchronization strategy substantially increases its operating cost. Performance therefore needs an explicit measurement and refinement process rather than an assumption that it will follow from successful translation.

\cat{Our approach.} We present \ourSystem, a multi-agent workflow that migrates a repository in two phases, each carried out by a team of specialized LLM agents. The \emph{translation phase} builds a target implementation that reproduces the required behavior of the source. The \emph{optimization phase} then improves the performance of that implementation. Agents perform the analysis, editing, and assessment, and hand work to one another by writing files rather than by conversing. A deterministic state machine decides which agent runs next and whether a phase may end, applying explicit rules to the outcomes those agents record. Termination is therefore determined by those rules, not by an agent's own judgment that its work is finished.

The translation phase never attempts the whole repository at once. It first analyzes the source, then divides the migration into an ordered sequence of \emph{milestones}, each one a slice of functionality together with the tests that must pass for it to count as done. Milestones are attempted one at a time, and each is checked against the tests of all earlier milestones as well as its own, so work on one slice cannot silently break another. Whether a slice is finished is judged by a fixed external oracle that the agents may neither author nor modify. 

Because a passing suite still cannot show that nothing was left behind, a \emph{parity verifier}---an LLM judge with no ability to edit code---then compares the components of the source against those of the target and reports whatever is missing or left as a placeholder. Crucially, a reported gap is not merely recorded. It is fed back into the plan as a new milestone, so the pipeline returns to work rather than finishing with a known omission, and the verifier runs again on the result.

The optimization phase begins once the translation is accepted. It alternates measurement against the project's benchmark with targeted improvement, and then runs the full test suite as a final milestone that can itself be repaired, so performance work cannot quietly trade behavior for speed. The two phases thus meet the three challenges in turn: milestones for scale, independent checks for self-validation, and performance measurement for efficiency.

Across nine subjects in three language pairs, \ourSystem achieves the best test pass rate on every subject against ReCodeAgent~\cite{recodeagent} and four simpler baselines that use the same model. It passes 1{,}166 of 1{,}180 Java-to-Python tests, all 63 C-to-Rust tests, and all 105 tests of a production daemon, against 1{,}029, 47, and 53 for ReCodeAgent. Five rounds of its optimization phase then improve five of the daemon's ten benchmarks by more than 5\%, cutting hardware reads by 21.6\% and idle CPU use by 12.2\%.

In summary, this paper makes the following contributions:

\begin{itemize}
\item \ourSystem, a two-phase repository-level migration workflow combining specialized agents with deterministic orchestration and explicit termination criteria.

\item A translation phase addressing scale and self-validation through ordered milestones, cumulative test gates, a fixed external oracle, and parity verification that converts detected gaps into new milestones.

\item An optimization phase alternating benchmark measurement and targeted improvement, followed by full-suite conformance testing and repair.

\item An evaluation on nine subjects in three language pairs, in which \ourSystem has the best pass rate on every subject and passes 18\% more tests than ReCodeAgent. On a production daemon, its optimization phase improves half of the benchmarks, by 25\% on average.
\end{itemize}

Section~\ref{sec:background} defines the problem. Section~\ref{sec:approach} describes the design. Sections~\ref{sec:evaldesign} and~\ref{sec:results} present the evaluation. Section~\ref{sec:related} reviews related work and Section~\ref{sec:conclusion} concludes.

\section{Problem Definition}
\label{sec:background}

We define repository-level migration as the discharge of a fixed set of \emph{obligations}. Each obligation is a requirement on the target repository, paired with the kind of evidence that can show it has been met. This section defines the inputs and outputs of a migration, the obligations they induce, and who may decide that an obligation has been discharged.

\cat{Migration instances.} A migration instance is a tuple
\begin{equation}
M = (S, L_t, \Sigma, \mathcal{O}, \mathcal{B}),
\label{eq:instance}
\end{equation}
where $S$ is the source repository and $L_t$ is the target language and environment. The \emph{scope} $\Sigma$ is the set of source components that the migration must reproduce, where a component is a module, type, function, or exported interface. Components outside $\Sigma$ are declared exclusions rather than gaps. The \emph{acceptance oracle} $\mathcal{O}$ is an executable test suite that is fixed before the migration begins and authored independently of it; we write $T \models \mathcal{O}$ when a target repository $T$ passes it. The optional benchmark suite $\mathcal{B}$ defines representative workloads and a cost measure $\mu(\mathcal{B}, T)$, such as time or peak memory.

A migration procedure applied to $M$ produces a target repository $T$, a suite $\mathcal{G}$ of tests that it authored itself, and a \emph{realization map}
\begin{equation}
R : \Sigma \rightharpoonup 2^{C(T)},
\label{eq:realization}
\end{equation}
which records, for each in-scope source component, the set of components of $T$ that realize it. $R$ is partial because a component may be omitted. It is set-valued because a migration may split one source component across several target components or merge several into one.

\cat{Obligations and acceptance.} Ideally, $T$ would be observationally equivalent to $S$ on every in-scope behavior. This condition cannot be checked in general, so we approximate it with three kinds of obligation, stated as the following conditions:
\begin{align}
&\textbf{C1 (build)} && T \text{ builds in the target environment;} \label{eq:c1}\\
&\textbf{C2 (behavior)} && T \models \mathcal{O}; \label{eq:c2}\\
&\textbf{C3 (coverage)} && R \text{ is total on } \Sigma \text{ and each } R(c) \nonumber\\
&&& \text{substantively implements } c. \label{eq:c3}
\end{align}
Each oracle test is a separate behavioral obligation, and each in-scope component is a separate coverage obligation. C3 rules out realizations that are present but vacuous, such as a stub with a compatible signature. C2 and C3 are independent of each other. A target can pass every oracle test while omitting a component that no test exercises, and a total $R$ says nothing about whether the realized components behave correctly. When a benchmark suite is given, the optimization objective is to reduce $\mu$ among the targets that satisfy C1--C3.

\cat{Progressive discharge.} For a repository of realistic size, no single step can discharge every obligation. A procedure therefore works in stages, and at any point each obligation is open, discharged, or deferred. A deferred obligation has been set aside, but its status is recorded so that it is not lost. This view leads to four requirements, each met by a mechanism described in Section~\ref{sec:approach}. First, each stage should take on a bounded subset of the obligations, which is the role of milestones. Second, work on one stage must not reopen obligations discharged in earlier stages, which cumulative test gates enforce. Third, a plan can miss obligations, so missed obligations must be found by inspecting the target and fed back into the plan; the parity loop provides this path. Fourth, explicit rules over the obligation state, rather than an agent's own judgment, should decide when a run ends, and the result should report which obligations remain open or deferred.

\cat{Who may discharge an obligation.} The tests in $\mathcal{G}$ are an output of the same procedure that produced $T$. As a result, $T \models \mathcal{G}$ can be achieved by weakening $\mathcal{G}$ just as easily as by improving $T$. The oracle $\mathcal{O}$, by contrast, is an input fixed before the run, so C2 cannot be satisfied in this way. For the same reason, coverage should be judged by an assessor that did not write $T$ and is not allowed to change it. Generated tests are still useful because they help localize defects during repair, but they do not count as acceptance evidence. Finally, a procedure's own stopping rule, which we call \emph{internal completion}, reports only that the procedure has finished. It does not show that C2 or C3 holds.

\section{\ourSystem Design}
\label{sec:approach}

\begin{figure*}[t]
\centering
\definecolor{oiBlue}{HTML}{0072B2}
\definecolor{oiSky}{HTML}{56B4E9}
\definecolor{oiOrange}{HTML}{E69F00}
\definecolor{oiVerm}{HTML}{D55E00}
\definecolor{oiGreen}{HTML}{009E73}
\begin{tikzpicture}[
  x=1cm, y=1cm,
  font=\sffamily\scriptsize,
  card/.style={line width=0.6pt, rounded corners=3pt, align=center,
    minimum width=2.05cm, minimum height=1.0cm, inner sep=1.5pt},
  plan/.style={card, draw=oiBlue, fill=oiSky!22},
  author/.style={card, draw=oiVerm, fill=oiOrange!26},
  assess/.style={card, draw=oiGreen!85!black, fill=oiGreen!15},
  book/.style={card, draw=black!55, dashed, fill=black!4, minimum width=1.15cm},
  input/.style={draw=black!60, line width=0.7pt, fill=black!8, rounded corners=3pt,
    align=center, inner sep=2.5pt},
  verdict/.style={draw=black, line width=0.9pt, fill=white, rounded corners=3pt,
    align=center, inner sep=2.5pt, minimum height=0.62cm},
  tab/.style={draw=black!45, fill=white, line width=0.4pt, rounded corners=1.5pt,
    inner sep=1.4pt, minimum height=0.34cm, anchor=north, align=center},
  band/.style={draw=black!70, line width=0.8pt, fill=white, rounded corners=4pt,
    minimum height=0.45cm, inner sep=0pt, anchor=west},
  panel/.style={draw=black!60, line width=0.9pt, fill=black!2.5, rounded corners=7pt},
  ptitle/.style={fill=black!75, text=white, font=\sffamily\footnotesize\bfseries,
    rounded corners=2pt, inner sep=3pt, anchor=north west, align=left},
  group/.style={draw=black!55, line width=0.6pt, fill=white, rounded corners=5pt},
  flow/.style={-{Stealth[length=1.8mm,width=1.5mm]}, line width=0.7pt},
  back/.style={flow, dashed, black!75},
  writes/.style={-{Stealth[length=1.7mm,width=1.4mm]}, line width=0.9pt, oiVerm},
  reads/.style={-{Stealth[length=1.6mm,width=1.3mm]}, line width=0.7pt,
    densely dotted, black!80},
  lbl/.style={fill=white, inner sep=0.8pt, font=\sffamily\scriptsize},
  badge/.style={circle, fill=white, inner sep=0.3pt, font=\normalsize},
]
\def\nm#1#2{{\footnotesize#1\,\textbf{#2}}}
\def\file{\faFile*[regular]\,}

\draw[panel] (1.95,0) rectangle (18.08,-5.0);
\node[ptitle] at (2.08,-0.12) {PHASE 1 \textperiodcentered{} \textmd{translation and assessment}};
\draw[panel, dashed] (6.3,-6.2) rectangle (18.08,-10.3);
\node[ptitle] at (6.42,-6.32) {PHASE 2\\ \textmd{optimization}\\ \textmd{(optional)}};
\draw[group] (8.36,-6.4) rectangle (13.12,-9.2);
\node[anchor=north east, align=right] at (13.04,-6.44)
  {\textbf{conformance milestone}\\ reuses the {\small\stepn{5}}\,--\,{\small\stepn{6}} repair loop};

\node[input, minimum width=1.75cm] (in) at (0.9,-2.7)
  {\textbf{Inputs}\\[3pt] {\small\faFolderOpen}\\ source repo\\[3pt] {\small\faSlidersH}\\ config, brief};

\node[plan]   (an) at (3.175,-2.7)  {\nm{\faMicroscope}{Analyzer}\\ design the target};
\node[plan]   (sc) at (5.605,-2.7)  {\nm{\faListOl}{Scoper}\\ order milestones};
\node[plan]   (pl) at (8.035,-2.7)  {\nm{\faProjectDiagram}{Planner}\\ plan + skeleton};
\node[book]   (se) at (10.015,-2.7) {\nm{\faCog}{select}\\ next $M_k$};
\node[author] (tr) at (11.995,-2.7) {\nm{\faCode}{Translator}\\ implement, repair};
\node[assess] (va) at (14.425,-2.7) {\nm{\faVial}{Validator}\\ unit + oracle[$G_k$]};
\node[assess] (pa) at (16.855,-2.7) {\nm{\faBalanceScale}{Parity}\\ source $\leftrightarrow$ target};

\node[tab] (anf) at (an.south) {\file analysis.md};
\node[tab] (scf) at (sc.south) {\file milestones.json};
\node[tab] (plf) at (pl.south) {\file plan.json};
\node[tab] (trf) at (tr.south) {\faCode\, code, unit tests};
\node[tab] (vaf) at (va.south) {\file report.json\\ \file skips.json};
\node[tab] (paf) at (pa.south) {\file parity.json};

\draw[flow] (in.east |- an.west) -- (an.west);
\draw[flow] (an) -- (sc);
\draw[flow] (sc) -- (pl);
\draw[flow] (pl) -- (se);
\draw[flow] (se) -- (tr);
\draw[flow] (tr) -- (va);
\draw[flow] (va) -- (pa);

\draw[back] ($(va.north)+(-0.4,0)$) -- ++(0,0.45) -| node[lbl,pos=0.25]{repair} ($(tr.north)+(0.4,0)$);
\draw[back] ($(va.north)+(0.35,0)$) -- ++(0,0.80) -| node[lbl,pos=0.25]{next milestone (passed or skipped)} ($(se.north)+(0.25,0)$);
\draw[back] ($(pa.north)+(-0.35,0)$) -- ++(0,1.15) -| node[lbl,pos=0.25]{retry deferred tests once} ($(se.north)+(-0.25,0)$);
\draw[back] ($(pa.north)+(0.35,0)$) -- ++(0,1.50) -| node[lbl,pos=0.3]{coverage gaps: append milestones} (sc.north);

\node[band, minimum width=7.6cm] (wc1) at (7.0,-4.45) {};
\node[anchor=west, font=\sffamily\scriptsize\bfseries] at (8.35,-4.45) {\small\faCode\ target working copy};
\draw[writes] (plf.south) -- (wc1.north -| plf.south);
\draw[writes] ($(trf.south)+(-0.35,0)$) -- ($(wc1.north -| tr)+(-0.35,0)$);
\draw[reads]  ($(wc1.north -| va)+(-0.35,0)$) -- ($(vaf.south)+(-0.35,0)$);

\node[verdict] (v1) at (16.855,-4.45) {\faFlagCheckered\ \textbf{phase-1 verdict}};
\draw[flow] (paf.south) -- (v1.north);

\node[input, minimum width=3.4cm, minimum height=0.8cm] (or) at (11.995,-5.6)
  {{\small\faVials}\,\faLock\ \textbf{acceptance oracle}\\ fixed input, never edited};

\node[assess] (be)  at (16.855,-8.15) {\nm{\faStopwatch}{Benchmarker}\\ measure speed};
\node[author] (op)  at (14.425,-8.15) {\nm{\faRocket}{Optimizer}\\ one change set};
\node[author] (trc) at (11.995,-8.15) {\nm{\faCode}{Translator}\\ repair mode};
\node[assess] (vac) at (9.565,-8.15)  {\nm{\faVial}{Validator}\\ unit + full oracle};
\node[verdict, minimum width=1.6cm, minimum height=0.9cm] (v2) at (7.3,-8.15)
  {\faFlagCheckered\ \textbf{final}\\ \textbf{verdict}};

\node[tab] (bef)  at (be.south)  {\file bench.json};
\node[tab] (opf)  at (op.south)  {\file optimize.json};
\node[tab] (trcf) at (trc.south) {\faCode\, code};
\node[tab] (vacf) at (vac.south) {\file report.json};

\draw[flow] (v1.south) -- (be.north);
\node[lbl, align=right, anchor=east] at (16.72,-5.6) {complete parity,\\ optimization on};
\draw[flow] (be) -- (op);
\draw[flow] (op) -- (trc);
\draw[flow] (trc) -- (vac);
\draw[flow] (vac) -- (v2);

\draw[back] ($(op.north)+(0.35,0)$) -- ++(0,0.45) -| node[lbl,pos=0.25]{next round} ($(be.north)+(-0.4,0)$);
\draw[back] ($(vac.north)+(0.4,0)$) -- ++(0,0.36) -| node[lbl,pos=0.25]{repair} ($(trc.north)+(-0.35,0)$);

\node[band, minimum width=11.4cm] (wc2) at (6.48,-9.8) {};
\node[anchor=west, font=\sffamily\scriptsize\bfseries] at (6.7,-9.8) {\small\faCode\ working copy};
\draw[reads]  ($(wc2.north -| vac)+(-0.35,0)$) -- ($(vacf.south)+(-0.35,0)$);
\draw[writes] ($(trcf.south)+(-0.35,0)$) -- ($(wc2.north -| trc)+(-0.35,0)$);
\draw[writes] ($(opf.south)+(-0.35,0)$) -- ($(wc2.north -| op)+(-0.35,0)$);
\draw[reads]  ($(wc2.north -| be)+(-0.35,0)$) -- ($(bef.south)+(-0.35,0)$);

\draw[reads, line width=0.9pt] (or.east) -- node[lbl, above]{gate $G_k$} (14.825,-5.6) -- (14.825,0 |- vaf.south);
\draw[reads, line width=0.9pt] (or.west) -- node[lbl, above]{full suite} (9.165,-5.6) -- (9.165,0 |- vac.north);

\foreach \n/\nd in {1/an, 2/sc, 3/pl, 4/se, 5/tr, 6/va, 7/pa, 8/be, 9/op}
  \node[badge] at (\nd.north west) {\stepn{\n}};
\node[badge] at (8.36,-6.4) {\stepn{10}};

\node[draw=black!40, line width=0.6pt, rounded corners=4pt, fill=white,
  minimum width=6.05cm, minimum height=4.1cm, anchor=north west] at (0,-6.2) {};
\node[anchor=north west, font=\sffamily\scriptsize\bfseries] at (0.12,-6.3) {Legend};
\foreach \x/\y/\sty/\txt in {0.25/-6.9/plan/plans, 3.1/-6.9/author/writes code,
                             0.25/-7.28/assess/assesses, 3.1/-7.28/book/no LLM}{
  \node[\sty, minimum width=0.42cm, minimum height=0.22cm, inner sep=0pt, anchor=west] at (\x,\y) {};
  \node[anchor=west] at (\x+0.47,\y) {\txt};}
\draw[flow]   (0.25,-7.78) -- ++(0.55,0) node[right, black]{next step};
\draw[back]   (3.1,-7.78)  -- ++(0.55,0) node[right, black]{loop (budget)};
\draw[writes] (0.25,-8.16)  -- ++(0.55,0) node[right, black]{writes code};
\draw[reads]  (3.1,-8.16)   -- ++(0.55,0) node[right, black]{runs code, tests};
\node[anchor=west] at (0.2,-8.64)  {\file file a step writes};
\node[anchor=west] at (3.05,-8.64) {\faCode\ working copy};
\node[anchor=west] at (0.2,-9.02)   {\faLock\ fixed input};
\node[anchor=west] at (3.05,-9.02)  {\faFlagCheckered\ verdict};
\node[anchor=west] at (0.2,-9.5)  {$G_k$\ \ oracle tests selected by $M_0,\dots,M_k$};
\node[anchor=west] at (0.2,-9.88)   {\faCogs\ \ a deterministic state machine takes every arrow};
\end{tikzpicture}
\caption{\ourSystem control flow and artifacts. Phase~1 (top) runs left to right; the optional Phase~2 (bottom, dashed) runs right to left after a complete parity verdict. Numbers mark the steps in the order traced in Section~\ref{sec:walkthrough}. The acceptance oracle between the panels is run by the Validator of each phase, on the cumulative gate $G_k$ in Phase~1 and in full in the conformance milestone, which reuses the Phase-1 Translator--Validator repair loop. Bands show the target working copy that agents write or run, and tabs name the file each step writes. Validation advances after either a pass or a recorded give-up, and eligible deferred-test retries take precedence over accepting the parity verdict. Terminal failure edges and bootstrap entry points are omitted.}
\label{fig:graph}
\end{figure*}
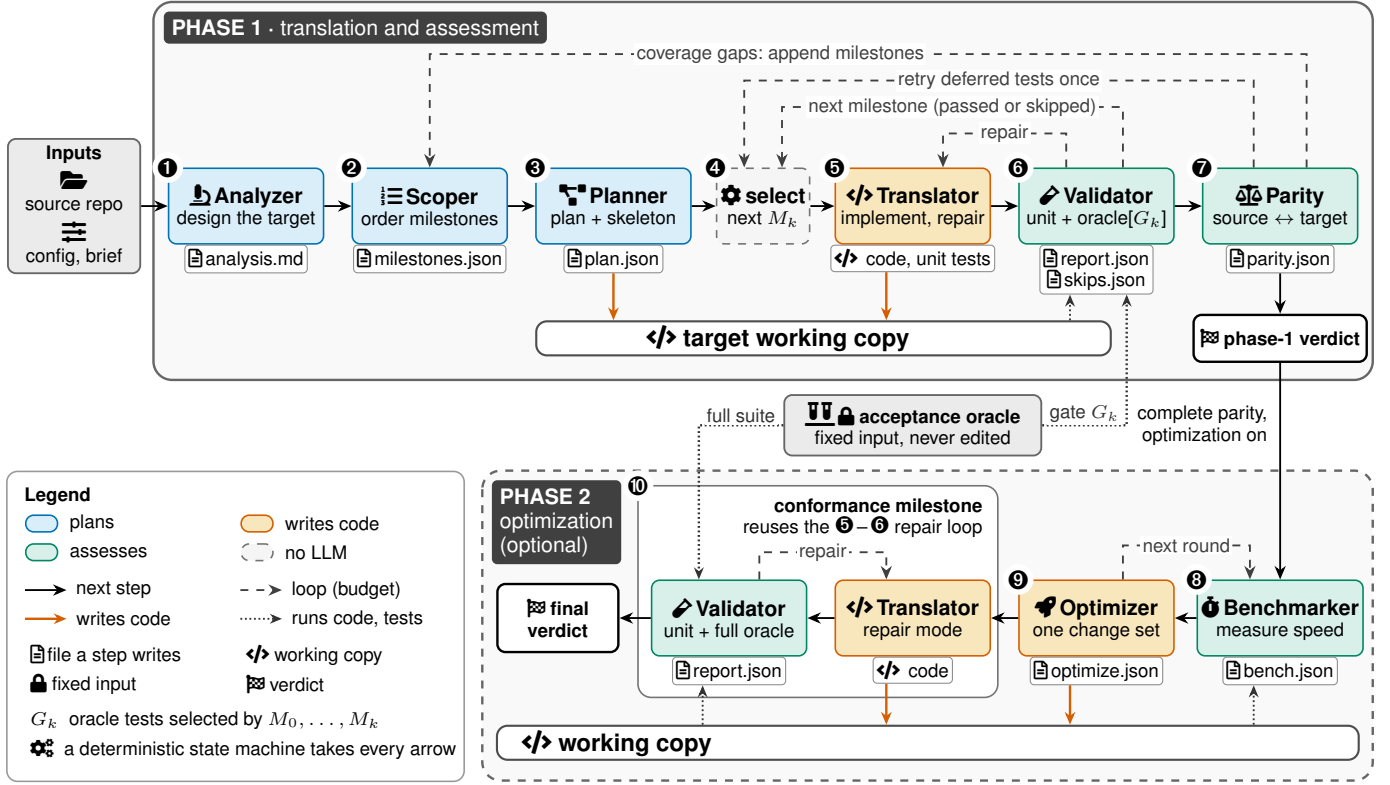

\ourSystem takes a migration instance (Section~\ref{sec:background}) and produces a target repository in two phases, shown in Figure~\ref{fig:graph}. \emph{Translation} constructs and assesses the port. \emph{Optimization} refines its performance using a supplied benchmark. In both phases the work is carried out by specialized LLM coding agents, each with a single responsibility, which hand work to one another by writing files. A deterministic state machine, not an agent, decides which agent runs next and when a phase ends. The remainder of this section is organized as follows. Sections~\ref{sec:control} and~\ref{sec:roles} describe the controller and the agent roles, Section~\ref{sec:oracle} defines the acceptance oracle against which behavior is judged, Section~\ref{sec:walkthrough} traces the end-to-end execution of a migration, and Section~\ref{sec:rationale} gives the rationale and remaining details of the mechanisms on which that execution relies.

\subsection{Separating reasoning from control}
\label{sec:control}

\ourSystem organizes migration as a sequence of evidence-producing actions. Its controller is a state machine that evaluates explicit predicates over persisted state to decide whether to advance, repair, defer, re-scope, or terminate. Actions launch a specialized LLM coding agent, run with a per-role profile (an \texttt{agent.md} file) that fixes the agent's standing instructions and the tools it may use, while each invocation adds a stage-specific prompt. Within its action, the harness runs its own loop of reading files, executing commands, and editing code. 

An LLM does not, however, choose the graph edge, so routing is deterministic given the observed state even though the agents' observations and edits are not. This prevents a conversational statement such as ``the migration is complete'' from replacing the configured stopping rule, and it exposes decisions that would otherwise be buried in a model's trajectory: repair budgets, outstanding skips, parity findings, and the reason for termination. The controller still relies on agent-produced reports, and deterministic processing of a mistaken report does not make it correct.

\subsection{Specialized roles and file handoffs}
\label{sec:roles}

Table~\ref{tab:agents} lists the eight roles and the artifact each produces.

\begin{table}[t]
\caption{Agent roles and authoritative handoff artifacts.}
\label{tab:agents}
\centering
\footnotesize
\begin{tabular}{@{}>{\raggedright\arraybackslash}p{1.6cm}>{\raggedright\arraybackslash}p{2.6cm}>{\raggedright\arraybackslash}p{3.5cm}@{}}
\toprule
\textbf{Role} & \textbf{Artifact} & \textbf{Responsibility} \\
\midrule
\multicolumn{3}{@{}l}{\emph{Translation phase}}\\
Analyzer & \texttt{analysis.md} & Source analysis and target design \\
Scoper & \texttt{milestones.json} & Initial work scope and gap-driven extensions \\
Planner & \texttt{plan.json};\newline skeleton & Implementation plan and target structure \\
Translator & Target working copy & Implementation, unit tests, and repairs \\
Validator & \texttt{report.json} & Test execution and failure diagnosis \\
Parity & \texttt{parity.json} & Source--target coverage assessment \\
\addlinespace
\multicolumn{3}{@{}l}{\emph{Optimization phase}}\\
Benchmarker & \texttt{bench.json} & Benchmark execution and measurements \\
Optimizer & \texttt{optimize.json};\newline code & Focused performance changes and rationale \\
\bottomrule
\end{tabular}
\end{table}

Agents exchange these files rather than relying on conversational handoffs. JSON reports carry machine-consumed verdicts; analysis and plans provide context for subsequent reasoning. Persistent artifacts make decisions easier to inspect and let an interrupted run recover its working context.

Only the Translator and the Optimizer change the target's implementation. The Planner contributes the skeleton, which holds stubbed signatures and test seams but no logic; the Analyzer and the Scoper write only their own artifacts; and the Validator, Parity agent, and Benchmarker assess code without changing it. The Parity agent and the Benchmarker are given no file-editing tool, whereas the Analyzer, Scoper, and Validator have one because each writes its own artifact, and are instructed to write nothing else. Separating assessment from authorship reduces role conflicts: the agent explaining a failed test should not silently weaken that test, and the agent measuring a change should not choose the outcome.

\subsection{The acceptance oracle}
\label{sec:oracle}

Behavioral acceptance in \ourSystem is judged against an \emph{acceptance oracle}, the executable end-to-end test suite that its validation step runs, fixed before the migration begins and authored independently of it. It instantiates the oracle $\mathcal{O}$ of Section~\ref{sec:background}; in a deployed system, it can be an existing end-to-end acceptance suite. During translation the oracle is run on subsets chosen by test selectors, and after optimization it is run in full (Section~\ref{sec:walkthrough}).

Because the oracle tests the target end to end, it runs in an environment that reproduces the conditions under which the target operates, including the services and devices it depends on. Unit tests may replace such dependencies with mocks; an acceptance test may not, since its purpose is to observe the target as it will be deployed. Passing the oracle therefore provides evidence for condition C2 that rests on no test written during the migration.
The oracle is fixed before migration, and agents are instructed not to edit or regenerate it. 

Only the Validator interacts with the oracle, and it writes no code. The agents that change the implementation receive only its report, which gives pass and fail counts and a diagnosis of missing or incorrect functionality, enough to direct a repair without disclosing what the tests check. Withholding the test source is intended to limit overfitting in the sense of Goodhart’s law~\cite{goodhart1984}. It reduces this risk rather than eliminating it, since runner output can contain assertion text and expected values, and the Validator’s diagnosis may relay them. We therefore treat oracle conformance as evidence for C2, not proof of equivalence, and rely on parity assessment for in-scope behavior the oracle does not exercise. The Validator also runs the unit tests that the Translator writes ($\mathcal{G}$ in Section~\ref{sec:background}), but these carry little weight as verification because they come from the same agent as the code.

\subsection{End-to-end execution}
\label{sec:walkthrough}

This subsection traces a migration through Figure~\ref{fig:graph}, from configuration to final verdict, and introduces each mechanism where execution first depends on it. Numbered markers such as~\stepn{1} refer to the corresponding steps in Figure~\ref{fig:graph}.

\cat{Inputs.} Project configuration supplies the source and target languages, repository paths, build and test commands, loop budgets, and a migration brief stating required interfaces, architectural constraints, and exclusions. This keeps the orchestration independent of a particular language pair, though a new deployment still needs toolchains, commands, and acceptance tests; configuration portability is not demonstrated effectiveness across arbitrary languages. Agents build the target in a configured working copy and are instructed never to modify the oracle or any provided starting code and scaffolding.

\cat{Preparation.} Three agents prepare the migration. The Analyzer~\stepn{1} studies the source, including its dependencies and its own unit tests, and records a design for the target in \texttt{analysis.md}, but writes no target code. The design maps source modules to target modules, names target counterparts for third-party libraries, plans unit tests against mocked boundaries, and inventories the behaviors of the source. The Scoper~\stepn{2} partitions these behaviors into an ordered list of \emph{milestones} in \texttt{milestones.json}; when the configuration already declares the milestones, this step records them without invoking an agent. The first milestone is a skeleton that only has to build and start, each later one adds a slice of functionality that depends only on earlier ones, and the last covers the full contract. Each milestone also lists the \emph{test selectors} it adds, which are names of oracle tests or test modules that a configured template renders into the test runner's syntax, such as a pytest \texttt{-k} expression. The Planner~\stepn{3} sets up the working copy from the provided starting code and turns the design into an executable plan in \texttt{plan.json}, which lists every unit to translate, maps each source name to its target counterpart, and records the steps of each milestone. It also generates a compilable skeleton of the target, with stubbed signatures and mock seams for unit tests but no logic.

\cat{The milestone loop.} The controller then selects the milestones one at a time~\stepn{4}. The Translator~\stepn{5} implements the current milestone in the working copy, replacing the stubs of its slice with real logic, translating the corresponding source unit tests, and adding unit tests for new code. The Validator~\stepn{6} then tests the result at two layers without changing any code (Section~\ref{sec:oracle}). It runs the unit tests, and it runs the oracle restricted to the milestone's \emph{cumulative gate}, the selectors of this milestone together with those of every earlier one, so that new work cannot silently break earlier work. The milestone passes only if both layers pass, and a build failure counts as a failure. The Validator records its verdict in \texttt{report.json}, with pass and fail counts for each layer and, for each failure, the likely cause and a repair hint. On failure, the Translator works from that report, determines whether each failure is a defect or functionality not yet translated, and fixes its root cause; the Validator then runs again, until the milestone passes or its attempt budget runs out. Under the default skip-on-give-up policy, a milestone that exhausts its budget is recorded as skipped, and its failing oracle tests are \emph{deferred}, which removes them from later gates but keeps them on record in \texttt{skips.json}. Without that policy, giving up ends the run as a failure. In every other case the controller advances to the next milestone, whether the current one passed or was skipped.

\cat{Coverage assessment.} Passing every gate does not show that nothing was left behind, because no test asserts that a component exists. After the last milestone, the Parity agent~\stepn{7} therefore compares source and target component by component, from modules and types to functions and the public interface, ignoring anything the brief excludes. Without editing either code base, it writes \texttt{parity.json}, which holds a completeness verdict and, for each missing, stubbed, or partial component, its source location and a suggested milestone. The controller then chooses among three paths, the first two bounded by a parity-round budget. If some deferred tests have not yet had their one retry, it appends a single milestone that re-enables them. Otherwise, if gaps remain, the Scoper~\stepn{2} appends milestones for them, the Planner~\stepn{3} revisits the plan, and the loop resumes at the new work. In both cases the Parity agent runs again once the appended milestones conclude. Otherwise the translation phase ends, successfully if parity reported complete coverage and unsuccessfully if gaps remain.

\cat{Optimization.} The second phase begins after a complete parity verdict. It requires optimization to be enabled, a positive round budget, and parity checking; without these prerequisites, execution ends after the translation phase. Before the first change, the controller snapshots the working copy so that the pre-optimization port can be recovered. In each round, the Benchmarker~\stepn{8} runs the benchmark command against the working copy and records the measurements in \texttt{bench.json} without editing anything. Without a configured command, the Benchmarker devises one in the first round for later rounds to reuse, though configuring one is preferable. The Optimizer~\stepn{9} then uses them, together with the history of earlier rounds, to make one small change set that preserves behavior. It records in \texttt{optimize.json} the files it touched, its rationale, and the effect it expects, and it is instructed to leave the unit tests passing. Changes accumulate from round to round. When the round budget is spent, the controller appends a final \emph{full-suite conformance milestone}~\stepn{10}, which runs the entire suite with no selector, deferred tests included, and repairs any failure through the Translator--Validator loop of steps~\stepn{5}--\stepn{6}. Its instructions are to restore behavior while keeping performance work where possible, and to undo an optimization that cannot be made correct rather than weaken a test. The run succeeds only if this milestone passes. Throughout, the controller persists its state, so an interrupted run can resume from its recorded progress.

\subsection{Design rationale and details}
\label{sec:rationale}

\cat{Milestones as regression boundaries.}\label{sec:milestones}%
Cumulative gates make each milestone a regression boundary without requiring the full suite at every intermediate step. Formally, let $\tau_k$ be the selectors introduced by milestone $k$. Its cumulative gate is
\begin{equation}
G_k = \operatorname{dedup}\!\left(\bigcup_{j=0}^{k} \tau_j\right),
\label{eq:gate}
\end{equation}
which the controller renders into the configured test-runner syntax. The boundary is, however, only as comprehensive as the selected tests, since accumulating selectors does not establish that every relevant test exists or was selected correctly. Each milestone's attempt budget counts every round of implementation and validation, including the first. When a milestone concludes, the controller clears its validation report from state, so that old failures are not mistaken for the new milestone's initial task.

\cat{The parity-completeness loop.}\label{sec:parity}%
The parity report estimates condition C3 of Section~\ref{sec:background}. Its gaps are the components of $\Sigma$ on which the Parity agent believes $R$ to be undefined or vacuous. Users can also confirm the structural half of C3, that $R$ is total on $\Sigma$, with deterministic tools such as a comparison of symbol inventories extracted from both code bases, so the agent's judgment is needed mainly to decide whether each counterpart is substantive, which no syntactic check can establish. 

The key design choice is to connect this assessment back to execution, which creates an explicit path from ``a source obligation is missing'' to implementation, validation, and another assessment. Existing milestone identifiers are intended to remain stable as gap milestones are appended. The loop complements, rather than replaces, earlier inventory and coverage techniques. A plan is a statement of intended work; a post-translation assessment asks whether that intent is reflected in the artifact. The loop is useful precisely when the initial plan was incomplete or an implementation step did not fulfill it. Its effectiveness nevertheless depends on the Parity agent recognizing the omission and the subsequent work repairing it; neither outcome follows from the graph structure alone.

\cat{Bounded repair and deferral.}\label{sec:giveup}%
One difficult milestone should not necessarily prevent useful work on the rest of the repository. When a milestone gives up, the controller attempts to identify its failing oracle tests from the Validator's report and defers only those, so the milestone's passing tests and all other obligations remain in later gates. This is a deliberate trade-off between making progress and retaining an unresolved acceptance obligation, not a claim that the skipped behavior is satisfactory. Starting a retry milestone re-enables its deferred tests and records that their dedicated retry has been used. If they fail again, they remain recorded but receive no further dedicated retry. Thus, the policy provides \emph{at most one} such retry per test, not an unconditional guarantee that every deferred test is eventually repaired.

\cat{Benchmark-guided optimization.}\label{sec:optimize}%
The performance phase separates the question of whether a change improves measured performance from the question of whether the resulting program still satisfies its behavioral contract. The Optimizer also consults the migration brief, and the history of earlier rounds is append-only, so every change and its rationale stay on record. Accumulating changes across rounds is deliberate. No round is validated against the full suite or rolled back on failure, and the controller does not enforce the Optimizer's instruction to keep the unit tests passing. The round budget, rather than a measured speedup threshold, governs the loop. This design aims to avoid paying for an expensive end-to-end run after every local optimization and to separate performance exploration from final behavioral repair. Since the conformance milestone uses the normal repair loop, a failure can be diagnosed and repaired rather than immediately discarding all accumulated changes. ``One milestone'' does not mean one test execution, since its bounded repair loop may execute the suite several times. The current schedule benchmarks before each optimization change, so it does not automatically remeasure the final change or any subsequent conformance repair. A performance evaluation must therefore benchmark the final accepted artifact separately. The phase supplies a refinement mechanism, not a guarantee of speedup or retention of the best-performing intermediate program.

\cat{Persistence and re-entry.}\label{sec:persistence}%
State is kept in a database under a run identifier, and the milestone list is reloaded from disk, so resuming preserves recorded progress, counters, and deferrals rather than reconstructing them from a transcript. Entry at a named milestone, at parity, or at benchmarking is supported when the prerequisite artifacts are supplied, which is useful for reassessment but constitutes operator-provided starting conditions rather than evidence that omitted stages succeeded. Persistence does not provide exactly-once execution. Filesystem edits and external commands are not atomic with a checkpoint, so replaying an interrupted action can encounter its earlier side effects. Reproducible routing is likewise distinct from reproducible generated code; replication requires preserving the source and target revisions, configuration, model and tool versions, artifacts, and environment, as Section~\ref{sec:evaldesign} does.

\section{Experimental Setup}
\label{sec:evaldesign}

We compare \ourSystem with ReCodeAgent and four simpler baselines that use the same model, on two benchmark suites and a production migration that also exercises the optimization phase. This section describes the implementation we evaluate, the subjects and their test suites, the baselines, and how we score their outputs.

\subsection{Implementation}
\label{sec:impl}

\ourSystem is implemented in Python on two existing systems. Its controller (Section~\ref{sec:control}) is an Apache Burr~\cite{burr} application in which every stage of Figure~\ref{fig:graph} is an action and every routing rule is a transition guarded by a predicate over the application state. The repair edge from validation back to translation, for example, is taken while the milestone has not passed and its attempt count is below the budget. Burr persists the state to an SQLite database after every action, keyed by the run identifier, which supports the resumption and re-entry of Section~\ref{sec:persistence}. An action that needs an agent launches one non-interactive GitHub Copilot CLI~\cite{copilotcli} process with the configured model and reasoning effort and with all permission prompts pre-approved, and Copilot CLI runs the agent loop of reading files, executing commands, and editing code. The controller logs the process's JSON event stream but never routes on it, so what an agent says in its session cannot change the route; only the files it writes can, such as the Validator's \texttt{report.json}.

The eight roles of Table~\ref{tab:agents} are Copilot custom agents, each defined by an \texttt{agent.md} profile. The profile's header names the role and lists the tools it may use, and its body states the role's standing instructions and hard boundaries; the Validator's, for instance, lets it run the two test commands and write its report but forbids editing the implementation, the tests, or the oracle. Copilot CLI enforces the tool list, whereas the boundaries are instructions only. Each invocation also receives a stage prompt rendered from a template. The templates encode the method and are shared across language pairs. The project configuration fills in the languages, paths, build and test commands, and migration brief, and the controller fills in the current milestone, whether it is to be implemented or repaired, the Validator's failures, and the cumulative gate. Each subject's test suite serves as the pipeline's acceptance oracle. 

\subsection{Subjects}
\label{sec:subjects}

We evaluate on nine subjects from three suites that differ in language pair and test provenance (Table~\ref{tab:subjects}), and we report each suite separately. Section~\ref{sec:overview} reports all three.

\begin{table}[t]
\caption{Subjects, grouped by suite.}
\label{tab:subjects}
\centering
\footnotesize
\setlength{\tabcolsep}{4pt}
\begin{tabular}{@{}lrrr@{}}
\toprule
\textbf{Subject} & \textbf{Source LOC} & \textbf{Modules} & \textbf{Tests} \\
\midrule
\multicolumn{4}{@{}l}{\emph{AlphaTrans, Java $\rightarrow$ Python}}\\
commons-cli        & 2{,}315 & 22 & 381 \\
commons-csv        & 2{,}308 & 11 & 298 \\
commons-fileupload & 1{,}634 & 30 & \phantom{0}39 \\
commons-validator  & 6{,}950 & 63 & 462 \\
\addlinespace
\multicolumn{4}{@{}l}{\emph{CRUST-Bench, C $\rightarrow$ Rust}}\\
cset                 & \phantom{0,}598 & \phantom{0}1 & \phantom{0}15 \\
c-aces               & \phantom{0,}777 & \phantom{0}7 & \phantom{0}11 \\
lambda-calculus-eval & 1{,}399 & \phantom{0}7 & \phantom{0}22 \\
inversion\_list      & \phantom{0,}695 & \phantom{0}1 & \phantom{0}15 \\
\addlinespace
\multicolumn{4}{@{}l}{\emph{SONiC xcvrd, Python $\rightarrow$ Rust}}\\
xcvrd daemon & 15{,}554 & 37 & 105 \\
\bottomrule
\end{tabular}
\end{table}

\begin{figure*}[t]
\centering
\includegraphics[width=\textwidth]{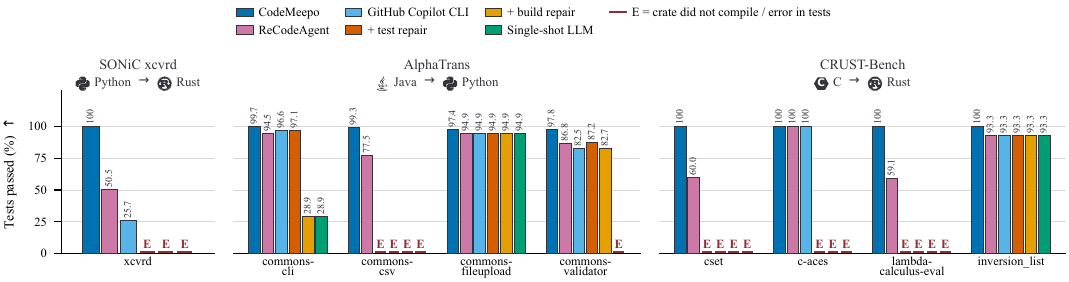}
\caption{Test pass rate of every arm on every subject, one panel per suite. \textbf{E} marks a crate that did not compile or a test suite that ended in an error, and is not a zero. \ourSystem achieves the best pass rate on every subject and passes 18\% more tests than ReCodeAgent using the same model.}
\label{fig:overview}
\end{figure*}

\cat{AlphaTrans (Java to Python).} This suite comprises the four Tier~A projects of AlphaTrans~\cite{alphatrans}, Apache Commons libraries for command-line parsing, CSV processing, file uploads, and input validation. Their tests are AlphaTrans's manually verified Python tests. Every arm receives the Java \texttt{src/main} tree.

\cat{CRUST-Bench (C to Rust).} CRUST-Bench~\cite{crustbench} pairs C projects with safe Rust interfaces and Rust tests. We take four of its projects (a hash set, a homomorphic encryption scheme, a lambda-calculus interpreter, and an integer-set library), chosen by assertion density because the largest projects carry almost no assertions. As a negative control we also score the benchmark's \texttt{unimplemented!()} stubs; one \texttt{inversion\_list} test passes on them and stays in the denominator.

\cat{SONiC xcvrd (Python to Rust).} Translation benchmarks are built mostly from small standalone projects whose tests run in isolation; CRUST-Bench's projects average about 958 lines, and earlier C-to-Rust benchmarks targeted single functions~\cite{crustbench}. Production migrations often replace components of deployed systems instead, so we add xcvrd, the transceiver daemon of the open-source network operating system SONiC~\cite{sonic}, which manages a switch's pluggable optical modules. Its concurrent worker threads drive per-module state machines, read and write three Redis databases, and call a vendor platform plugin that the Rust port must keep reaching through a bridge. The daemon also depends heavily on the rest of SONiC, building on its shared libraries and exchanging state with other services, and a port must preserve all of these interactions. It is a plausible rewrite target as well, since developers adopt Rust largely for its performance and safety~\cite{fulton2021rust} and xcvrd's polling cost grows with the number of ports.

Because the daemon's behavior is observable only through the system around it, we wrote an end-to-end suite of 105 tests that treats the daemon as a black box. It runs on a SONiC virtual switch with emulated optical modules, drives the daemon through the channels a deployment would use (plugging and unplugging modules, writing their EEPROM, injecting faults, and changing port configuration), and observes the published tables together with a trace of every EEPROM read and write. The tests cover hot-plug handling, decoding for two module standards, the CMIS bring-up state machine, fault recovery, and restarts, and three conformance tests compare the published tables with snapshots from the upstream Python daemon, which serves as the reference. Each session first flushes the daemon's tables and requires it to repopulate them, so stale rows cannot produce a pass, and on an earlier version of the suite a stand-in daemon that wrote placeholder rows failed every test that depended on the daemon. Because the suite never imports the daemon, it grades the Python original and the Rust port identically.

\subsection{Baselines}
\label{sec:arms}
We compare \ourSystem with ReCodeAgent, two unscaffolded baselines, and two repair loops applied to the single-shot output. All suite runs use \texttt{claude-sonnet-5} at medium reasoning effort, so differences reflect scaffolding rather than model choice, and \ourSystem runs with \texttt{max\_iter}~$=6$ and \texttt{max\_parity\_rounds}~$=4$. 
\begin{itemize}
\item \textbf{GitHub Copilot CLI.} A single software engineering agent in the client that hosts our agents (Section~\ref{sec:impl}), with all tools enabled and all permissions granted, given the translation goal, the whole repository, and the command that runs the subject’s suite in one prompt as in CRUST-Bench~\cite{crustbench}, but lack \ourSystem's workflow structure: no milestones, parity loop, or repair.

\item \textbf{Single-shot LLM.} One independent direct-API request per module, matching AlphaTrans's class-by-class ablation~\cite{alphatrans}, with no shared state, retries, or feedback.
\item \textbf{+ build repair, + test repair.} Up to three repair iterations on the frozen single-shot output, guided by compiler or parse-and-import diagnostics, the signal of \ourSystem's \texttt{build\_check}, and for test repair also by the suite's failure output. Following CRUST-Bench~\cite{crustbench}, both keep the final iteration rather than the best one.
\item \textbf{ReCodeAgent~\cite{recodeagent}.} A state-of-the-art multi-agent workflow whose agents analyze, plan, translate, and validate a repository. We implemented ReCodeAgent with Copilot CLI custom agents orchestrated in Burr.

\end{itemize}

All arms operate under the same information conditions. Every arm except the single-shot LLM, which only issues one request per module, can execute the subject’s suite and read its output, and none is given the test source. What differs is whether a workflow is designed to use that signal. This mirrors practice when an organization migrating a deployed system usually already owns an acceptance suite, and the question we study is which workflow best turns that existing contract into a conforming port. \ourSystem is designed to treat the suite as its acceptance oracle at every milestone gate and test repair reads its failure output, whereas the other baselines do not systematically integrate this feedback into their workflows. Using external acceptance evidence to drive migration is a central capability of \ourSystem. Our comparison evaluates the resulting end-to-end workflows.

Scores are recomputed independently from frozen artifacts, not read from any pipeline's self-report. Scoring stages the candidate source together with a pristine copy of the test suite into a throwaway tree, so no arm can have modified the tests it is judged by. We report passed out of passed $+$ failed.

\section{Results}
\label{sec:results}

\subsection{Comparison with baselines}
\label{sec:overview}

\emph{\ourSystem achieves the best pass rate on every subject, is matched only where it passes every test, and leads by the widest margin on the production xcvrd daemon.} Figure~\ref{fig:overview} places every arm on every subject. Only \ourSystem and ReCodeAgent produce a scoreable artifact on all nine, and \ourSystem passes more tests than ReCodeAgent on eight of them, while both pass every test on \mbox{c-aces}, 1{,}166 against 1{,}029 of AlphaTrans's 1{,}180 tests, 63 against 47 of CRUST-Bench's 63, and 105 against 53 of xcvrd's 105. Overall, it passes 18\% more tests than ReCodeAgent, 1{,}334 against 1{,}129.

\cat{Cross-module agreement.} The single-shot baseline breaks wherever modules must agree with one another. On commons-cli, 243 of its 271 failures reduce to one naming disagreement between callers and the interface skeleton, a similar disagreement makes the single-shot commons-csv uncollectable, neither unscaffolded baseline produces a collectable commons-csv, and in Rust the same defect stops three of the four single-shot crates from compiling. The damage follows coupling rather than size, since every arm scores within one test of every other on commons-fileupload, whose dependency graph is the flattest, and the one single-shot crate that compiles, \texttt{inversion\_list}, has no cross-module surface.

The Copilot CLI baseline holds the whole repository in view and largely agrees with itself, reaching 368 and 381 on the two AlphaTrans subjects where the single-shot arm collapses. It still fails \mbox{cset} like every unscaffolded arm, because the C source relies on aliasing that Rust rejects and the ownership structure has to be reworked. Although its tools let it run the compiler, the agent ended its session without finishing the subjects it failed, leaving \mbox{cset} and \mbox{lambda-calculus-eval} uncompilable and commons-csv uncollectable.

\cat{Repair after the fact.} Repair recovers only part of the gap. Build-guided repair fixes import-time defects cheaply, turning commons-validator's collection errors into 382 of 462 passes with a single call, but it cannot see the runtime errors that dominate commons-cli. Test-guided repair reads the test failures and reaches 370 and 403 on those two subjects, still below \ourSystem's 380 and 452, and no repaired crate compiles in CRUST-Bench.

ReCodeAgent is the strongest baseline in every suite and the only other arm with a scoreable artifact everywhere, yet it trails \ourSystem by 65 tests on commons-csv, 51 on commons-validator, and 52 on xcvrd. Three design differences separate it from \ourSystem. ReCodeAgent translates the repository as a whole, with no milestones that break the work into gated slices, which leaves its final port incomplete. Its Validator, though able to run the subject's suite, checks the port mainly against tests it translates or writes itself, which scales poorly to large repositories and can miss behavior. And it has no parity check that compares the finished port with the source and turns each gap into new work, so nothing confirms that everything in scope was translated.

\cat{The production subject.} The gap is widest on xcvrd, whose behavior depends on databases, a platform plugin, and concurrent tasks. The Copilot CLI passes 27 of its 105 tests and ReCodeAgent 53, and the single-shot output and both repair loops produce nothing the suite can score, while \ourSystem passes all 105, including the tests it deferred during the run. Its parity loop also recovered four source behaviors that no test exercises and added each as a new milestone, which no pass rate in Figure~\ref{fig:overview} can show.

\subsection{\ourSystem on xcvrd}
\label{sec:xcvrd-run}

After the five optimization rounds of phase~2, the optimized port beats the port it started from by more than 5\% on five of the ten benchmarks, and it still passes every test.

\begin{figure}[t]
\centering
\includegraphics[width=\columnwidth]{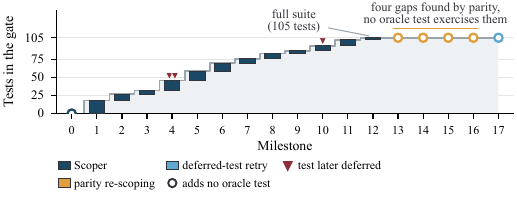}
\caption{Oracle tests gated at each phase-1 milestone of the xcvrd run.}
\vspace{-10pt}
\label{fig:abl-translation}
\end{figure}

\begin{table}[t]
\caption{The last recorded session of each agent role in the xcvrd run.}
\label{tab:sessions}
\centering
\footnotesize
\setlength{\tabcolsep}{3pt}
\begin{tabular}{@{}lcrrl@{}}
\toprule
\textbf{Role (unit of work)} & \textbf{Ph.} & \textbf{Model} & \textbf{Calls} & \textbf{Wrote} \\
\midrule
\rowcolor{black!6}
Analyzer               & 1 & 49\%  & 66  & analysis document \\
Scoper (re-scope)      & 1 & 96\%  & 9   & milestone list \\
\rowcolor{black!6}
Planner                & 1 & 98\%  & 53  & plan \\
Translator (M17)       & 1 & 18\%  & 163 & 10 crate files \\
\rowcolor{black!6}
Validator (M16)        & 1 & 3.5\% & 16  & its report \\
Parity agent           & 1 & 96\%  & 60  & its report \\
\rowcolor{black!6}
Benchmarker            & 2 & 5.9\% & 12  & --- \\
Optimizer (last round) & 2 & 97\%  & 40  & its round record \\
\rowcolor{black!6}
Validator (last round) & 2 & 5.3\% & 13  & its report \\
\bottomrule
\end{tabular}
\end{table}

\cat{One run, step by step.} Figure~\ref{fig:abl-translation} follows phase~1, the translation phase, milestone by milestone. The Analyzer, Scoper, and Planner turned the 37 source modules into 73 translation units scheduled over thirteen milestones, M0 to M12, and the cumulative gate grew with each of them until M12 ran the full 105-test suite. The Parity agent then found four source behaviors that no oracle test exercises. Re-scoping appended the first three as M13--M15, debouncing a newly inserted module, holding start-up until the port configuration is complete, and keeping the DOM polling set in step with ports added and removed at run time. A further parity round added the original's command-line options as M16, and M17 gave the three tests deferred along the way their single retry, after which the Parity agent reported complete coverage. Four of the eighteen milestones thus add behavior that no pass rate can show, and by the end of phase~1 the port carried 320 unit tests, up from 25 at M0.

Table~\ref{tab:sessions} shows how the work divided among the roles. Its Model column is the share of each session spent in model calls; the rest went to the tool calls counted in its Calls column, such as reading files, building the crate, and running the unit tests and the oracle. The judging roles never wrote to the crate. The Validators and the Benchmarker, which run the tests and benchmarks, spent only 3.5--5.9\% of their sessions in model calls, whereas the Scoper, Planner, Parity agent, and Optimizer spent 96--98\% of theirs there, so the oracle, not the model, sets the pace of a milestone.

\begin{figure}[t]
\centering
\includegraphics[width=\columnwidth]{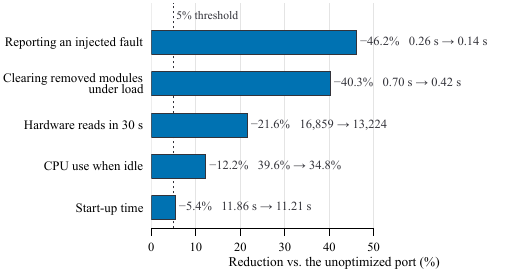}
\caption{Benchmarks on which the optimized port improved by more than 5\% over the unoptimized port, from one back-to-back run of each. Labels give the improvement and the values before and after.}
\vspace{-15pt}
\label{fig:opt-gains}
\end{figure}

\cat{Optimization.} None of the baselines optimizes performance, so we compare \ourSystem's port only with the unoptimized port it started from. We ran the optimization loop of phase~2 for five rounds, and in each the Optimizer proposed one change set.
The full-suite conformance milestone then ran all 105 tests on the accumulated changes, repairing any failure or undoing a change that could not be made correct, so the final port retains only optimizations that pass the full oracle suite. We then benchmarked both versions on the same machine, with ten benchmarks covering start-up, reactions to hardware events, idle resource use, steady-state work, and shutdown, 23 measurements in all. Counting a change of more than 5\% as an improvement, these five rounds improved five of the ten benchmarks, by 25\% on average (Figure~\ref{fig:opt-gains}). The optimized port reports an injected hardware fault 1.9 times as fast, clears removed modules 40.3\% faster during a burst of insertions and removals, issues 21.6\% fewer hardware reads, uses 12.2\% less CPU when idle, and finishes start-up 5.4\% sooner. The rest of the benchmarks changed by less than 5\%.

\section{Related Work}
\label{sec:related}

\cat{Language translation and repository evolution.} Neural code translation began at function granularity~\cite{transcoder}, later grounded by executing unit tests to filter unreliable outputs~\cite{transcoderst}, and empirical study shows that syntactically valid translations can still be semantically defective~\cite{lostintranslation}. Repository-level migration adds dependencies among translated units and requires reconstructing the surrounding project. AlphaTrans addresses this through compositional decomposition, target skeletons, incremental translation, and validation using source and translated tests~\cite{alphatrans}; we evaluate on four of its projects. Zhang et al.\ translate Go projects to Rust in validated fragments~\cite{zhang2025validated}. CodePlan addresses the related but distinct problem of propagating repository-wide edits from an initial change specification~\cite{codeplan}. Other work eases long or domain-specific inputs by chunking~\cite{chakaravarthy2026long} or summary augmentation~\cite{bhardwaj2026cobol}. These improve the information available to a translation step; \ourSystem operates at the workflow level, where milestones define integration boundaries and post-translation findings can create work after the original plan is exhausted.

\cat{Multi-agent migration.} CoMRA~\cite{comra2026} combines retrieval-augmented generation and multi-agent collaboration for Qt4-to-Qt5 version migration, emphasizing migration knowledge rather than the control rules \ourSystem relies on. ReCodeAgent~\cite{recodeagent} combines analysis, planning, translation, and validation for language-agnostic repository migration, and it already describes scripts checking fragment existence and file omission as well as comparisons between exercised tests and planned functions. Unlike \ourSystem, it translates the repository as a whole without gated milestones, validates mainly against tests it translates or writes itself, and has no parity check that turns each gap into new work. Our design emphasizes a dedicated post-translation parity assessment whose findings feed a bounded re-scoping loop, integrated with cumulative gates and explicit deferred-test retry. Project-level C-to-Rust agents orchestrate translation along dependencies~\cite{orbit2026} or repair a port where its regenerated documentation disagrees with the source's~\cite{rustprint2026}. ChatDev~\cite{chatdev}, MetaGPT~\cite{metagpt}, and StateFlow~\cite{stateflow} coordinate agents through dialogue, operating procedures, or state machines; \ourSystem routes on rules, never on model decisions.

\cat{Feedback and behavioral validation.} Compiler feedback can improve compilability without establishing behavioral correctness~\cite{kjellberg2026compiler}, while Weiss et al.~\cite{weiss2026feedback} study C-to-Rust feedback loops using compiler diagnostics and differential testing. Our repair results (Section~\ref{sec:overview}) sharpen this, since on a dynamically typed target a parse-and-import check cannot observe the runtime attribute errors that dominate a badly assembled translation. Models can self-debug from execution feedback~\cite{selfdebug}, and ACToR fuzzes for divergence from the C source~\cite{actor2025}. Other approaches exploit stronger structure in particular language pairs. VERT~\cite{vert} model-checks candidate Rust against a reference, SACTOR~\cite{sactor} combines static analysis with FFI-based verification, and Syzygy~\cite{syzygy} uses dynamic analysis in joint code and test translation. These should not be conflated with LLM-based parity inspection, because \ourSystem's coverage report is a fallible assessment, not a verifier.

\cat{Performance optimization.} LLMs can learn performance-improving edits~\cite{pie}, be guided by evolutionary search~\cite{sbllm}, or revise code against execution profiles~\cite{effilearner}, yet trail experts on real repositories~\cite{sweperf}. \ourSystem optimizes the port it just produced and returns all accumulated changes to the conformance milestone.

\cat{Benchmarks and agentic software engineering.} CRUST-Bench~\cite{crustbench} supplies safe Rust interfaces and tests for C projects, RepoTransBench~\cite{repotransbench} supports executable repository-level assessment, and RepoMod-Bench~\cite{repomodbench} hides its tests from the agents; the provenance of such contracts affects what success establishes, which is why \ourSystem separates generated tests from its oracle. The broader agentic literature contributes repository-task benchmarks~\cite{swebench}, agent--computer interfaces~\cite{sweagent}, and program-improvement workflows~\cite{autocoderover}, while Agentless~\cite{agentless} shows why workflow design should be evaluated rather than equated with autonomy. Agentic app generation~\cite{appbuild2026} and build repair~\cite{macho2026build} show the value of environment scaffolding and executable feedback. \ourSystem applies these principles to behavior-preserving migration; its distinction is the specific integration of progress accounting with post-translation coverage feedback, not a claim that deterministic orchestration, multiple agents, or repair loops are individually new.

\section{Conclusion}
\label{sec:conclusion}

Repository-level migration must split work that exceeds one model context into milestones without breaking earlier ones, accept a port only on evidence that its agents did not author, and keep a behaviorally correct port from being slow.

\ourSystem meets these challenges with explicit workflow structure, in which deterministic routing coordinates eight specialized roles, cumulative milestone gates support regression checking, parity findings reopen the plan, bounded repair records unresolved obligations, and accumulated performance changes return to a repairable conformance milestone. These mechanisms make progress and acceptance evidence inspectable without treating an LLM verdict as a proof.

Across nine subjects, \ourSystem achieves the best test pass rate on every subject. On a production daemon it passes all 105 tests against ReCodeAgent's 53, its parity loop added four milestones beyond the plan, and five optimization rounds improved half of the benchmarks, by 25\% on average.

\bibliographystyle{IEEEtran}
\bibliography{references}

\end{document}